\documentclass[12pt,aps,prb,preprint]{revtex4}
\usepackage[active]{srcltx}
\usepackage{graphicx}
\usepackage{epsfig}
\newcommand{\eref}[1]{(\ref{#1})}
\newcommand{\na}{\mbox{\boldmath $\nabla$}}

\newcommand{\T}{\mbox{\boldmath $T$}}

\newcommand{\ka}{\mbox{\boldmath $\kappa$}}
\newcommand{\ps}{\mbox{\boldmath $\psi$}}

\newcommand{\e}{\mbox{\boldmath $e$}}

\newcommand{\n}{\mbox{\boldmath $n$}}
\newcommand{\uu}{\mbox{\boldmath $u$}}

\newcommand{\ta}{\mbox{\boldmath $\tau$}}

\newcommand{\rr}{\mbox{\boldmath $r$}}

\newcommand{\J}{\mbox{\boldmath $J$}}

\newcommand{\B}{\mbox{\boldmath $B$}}

\newcommand{\kk}{\mbox{\boldmath $k$}}

\newcommand{\A}{\mbox{\boldmath$A$}}

\newcommand\fr{\displaystyle\frac}

\newcommand{\htts}{\mbox{\boldmath$\hat{t}\kern1pt$}}

\newcommand\lt{\left}
\newcommand\rt{\right}

\begin{document}
\title{A proposal of a UCN experiment to check an earthquake waves model}
\author{Vladimir K.Ignatovich}
 \affiliation{Frank Laboratory of Neutron Physics, Joint Institute
 for Nuclear research, Dubna, Moscow region 141980, Russia}
 \email{ignatovi@nf.jinr.ru}
\date{\today}

\begin{abstract}
Elastic waves with transverse polarization inside incidence plane
can create longitudinal surface wave (LSW) after reflection from a
free surface. At a critical incidence angle this LSW accumulates
energy density, which can be orders of magnitude higher than
energy density of the incident transverse wave. A specially
arranged vessel for storage of ultracold neutrons (UCN) can be
used to verify this effect.
\end{abstract}
\maketitle

\section{Introduction}

It is known~\cite{leu,fesh,die,tru,mus,fed} that reflection from
an interface of an elastic wave of a given mode in isotropic and
anisotropic media is accompanied with mode conversion.
In~\cite{ifan} it was noticed that at grazing angles smaller than
some critical angle the mode conversion can result in creation of
a longitudinal surface wave (LSW), which is alike to those
appearing at earthquakes. The most prominent property of LSW is
that critical angle it can accumulate an energy density orders of
magnitude larger that that of the incident plain wave.

In the next section the theory of elastic waves in isotropic media
is presented according to~\cite{ifan}, and in the third section an
experiment with ultracold neutrons (UCN) (see, for
example~\cite{ign}) is proposed to observe effect of LSW on UCN
storage.

\section{Elastic waves in isotropic media}

The main element of the theory of elastic waves\cite{land} is a
displacement vector $\uu(\rr,t)$ of a material point at a position
$\rr$ at a time $t$. Its Cartesian coordinates $u_i(\rr,t)$ obey
the Newtonian equation of motion
\begin{equation}\label{in}
\rho\fr{\partial^2}{\partial t^2}u_i(\rr,t)=\fr{\partial}{\partial
x_j}\sigma_{ij}(\rr,t),
\end{equation}
where $\rho$ is the mass density, $x_j$ are components of the
position vector $\rr$, $\sigma_{ij}$ is a stress tensor
\begin{equation}\label{in2}
\sigma_{ij}=c_{ijkl}u_{kl},
\end{equation}
which is proportional to the deformation or strain tensor $u_{ij}$
\begin{equation}\label{in1}
u_{jk}=\fr12\left(\fr{\partial u_j}{\partial x_k}+\fr{\partial
u_k}{\partial x_j}\right),
\end{equation}
and coefficients of proportionality $c_{ijkl}$ in \eref{in2} form
a tensor which in isotropic media looks
\begin{equation}\label{in4}
c_{ijkl}=\lambda\delta_{ij}\delta_{kl}+\mu(\delta_{il}\delta_{kj}+\delta_{ik}\delta_{lj}),
\end{equation}
where $\delta_{ij}$ is the Kronecker delta function, and $\lambda$
and $\mu$ are Lam\'e elastic constants.

With the stress tensor \eref{in2}, \eref{in4} the Newtonian
equation of motion (\ref{in}) for the displacement vector becomes
\begin{equation}
\label{eq5} \rho \ddot {u}_i = \nabla _j \sigma_{ij} = \mu [\Delta
u_i + \nabla _i (\nabla\cdot\uu)] + \lambda \nabla _i
(\na\cdot\uu).
\end{equation}

Usually the displacement vector $\uu(\rr,t)$ is represented as a
sum of two parts $\uu(\rr,t)=\na\varphi+\na\times\ps$, where
$\varphi$ is a scalar and $\ps$ is a vector potentials. This
tradition was broken in~~\cite{ifan}, and it helped to find new
unexpected effects. The solution of (\ref{eq5}) is represented in
the form of a complex plane wave
\begin{equation}\label{uu1}
\uu(\rr,t)=u_0\A\exp(i\kk\rr-i\omega t),
\end{equation}
where vector $\A$ ia a unit polarization vector, and $u_0$ is the
wave amplitude with dimension of length. After substitution of
$\uu$ from \eref{uu1} into (\ref{eq5}) one obtains an equation for
$\A$:
\begin{equation}
\label{eq7} \rho \omega ^2\A = \mu k^2\A + (\lambda + \mu
)\kk(\kk\cdot\A).
\end{equation}
Because of the linearity of the equation the amplitude $u_0$ of
\eref{uu1} does not matter and for some time will be omitted.

For a given propagation direction $\ka=\kk/k$ one can introduce
two orthonormal vectors $\e^{(1)}$ and $\e^{(2)}$, which are
perpendicular to $\ka$, and the orthonormal basis $\e^{(1)}$,
$\e^{(2)}$, $\e^{(3)}\equiv\ka$, in which the polarization vector
$\A$ looks
\begin{equation}\label{eq8}
\A=\alpha^{(1)}\e^{(1)}+\alpha^{(2)}\e^{(2)}+\alpha^{(3)}\e^{(3)}.
\end{equation}
After multiplication of (\ref{eq7}) by each of the $\e^{(i)}$ one
obtains three equations for the coordinates:
\begin{equation}\label{eq9}
(\rho \omega ^2-\mu k^2)\alpha^{(1,2)}=0,\qquad (\rho \omega
^2-[\lambda+2\mu] k^2)\alpha^{(3)}=0.
\end{equation}
They are independent and give three solutions: two for shear modes
$\A^{1,2}=\e^{(1,2)}$ with speed $c_t=\sqrt{\mu/\rho}$ and wave
number $k^{(1,2)}=\omega/c_t$, and one longitudinal mode
$\A^{3}=\ka$ with speed $c_l=\sqrt{(\lambda+2\mu)/\rho}$ and wave
number $k^{3}=\omega/c_l$.

Consider now reflection from a free surface at $z=0$ of the wave
\eref{uu1} propagating in the medium at $z<0$. In this case it is
convenient to choose basis vectors $\e^{(1,2)}\perp\ka$ in such a
way, that $\e^{(1)}$ is perpendicular to the incidence plain, and
$\e^{(2)}$ lies in it. The most interesting is reflection of the
shear mode $\A^{(2)}=\e^{(2)}$. The displacement vector, when the
incident wave is of this mode, looks
\begin{equation}\label{eq10}
\uu(\rr,t)=\exp(i\kk_\|\rr_\|-i\omega
t)\left[\A^{(2)}e^{ik^{(2)}_\bot
z}+r^{(22)}\A^{(2)}_Re^{-ik^{(2)}_{\bot}z}+
r^{(32)}\A^{(3)}_Re^{-ik^{(3)}_{\bot}z}\right],
\end{equation}
where $r^{(j2)}$ are reflection amplitudes of the mode 2 with
transformation into mode $j=2,3$, and
$k_\bot^{(j)}=\sqrt{k^{(j)2}-\kk_\|^2}$. It is important to notice
that $k_\bot^{(3)}<k_\bot^{(2)}$, therefore reflected longitudinal
wave propagates in nonspecular direction. Its grazing angle
$\varphi^{(3)}$ is less than $\varphi^{(2)}$ of the incident wave.
Therefore at some critical angle
$\varphi_c^{(2)}=\arccos(c_t/c_l)$ the longitudinal wave starts to
propagate along the surface, and at
$\varphi^{(2)}<\varphi_c^{(2)}$ the component $k_\bot^{(3)}$
becomes imaginary, i.e. the longitudinal reflected wave becomes
LSW. This is the most interesting phenomenon which, is a model of
the earthquake waves, because the most devastating effect of
earthquakes results from vibrations along the surface.

Let's find the energy of the LSW~\cite{kis}. To do that we need to
know reflection amplitudes. They are found from the boundary
condition, which requires the stress vector $\T$ with components
$T_j=\sigma_{jl}n_l$, where $n_j$ are components of a unit vector
$\n$ along the normal, to be zero at $z=0$. For stress tensor
\eref{in2}, \eref{in4} the stress vector for the displacement
$\uu(\rr,t)$ is
\begin{equation}
\label{eq13}\T\Big(\uu(\rr,t)\Big)=\lambda
\n(\na\cdot\uu)+\mu[\na(\uu\cdot\n)+(\n\cdot\na)\uu].
\end{equation}
Substitution of \eref{eq10} in it gives boundary condition in the
form
\begin{equation}\label{eq12a}
\B^{(2)}+r^{(22)}\B^{(2)}_{r}+r^{(32)}\B^{(3)}_{r}=0,
\end{equation}
where for every plane wave $\A\exp(i\kk\rr)$ the vector $\B$ is
defined as
\begin{equation}\label{eq12b}
\B=\lambda \n(\kk\cdot\A)+\mu[\kk(\A\cdot\n)+\A(\n\cdot\kk)],
\end{equation}
$\B^{(2)}$ is related to the incident wave with
$\kk^{(2)}=\kk_\|+\n k_\bot^{(2)}$, and $\B^{(j)}_r$ ($j=2,3$) are
related to the reflected waves with $\kk^{(j)}=\kk_\|-\n
k_\bot^{(j)}$.

Multiplication of \eref{eq12a} by two unit vectors: $\n$ and
$\ta=\kk_\|/|\kk_\||$ gives a system of two equations for
determination of $r^{(j2)}$. Their solution is
\begin{equation}\label{q5}
r^{(32)}=\fr{k^{(3)}}{k^{(2)}}\,
\fr{4k^{(2)}_{\bot}k_\|(k^{(2)2}-2k_\|^2)}{4k^{(3)}_{\bot}k^{(2)}_{\bot}k_\|^2+(k^{(2)2}-2k_{\|}^2)^2},\qquad
r^{(22)}=\fr{4k^{(3)}_{\bot}k^{(2)}_{\bot}k_\|^2-(k^{(2)2}-2k_{\|}^2)^2}{4k^{(3)}_{\bot}k^{(2)}_{\bot}k_\|^2+(k^{(2)2}-2k_{\|}^2)^2},
\end{equation}
and it is seen that imaginary $k^{(3)}_{\bot}$ makes
$|r^{(22)}|=1$, which means the total reflection of the mode 2.

Because of energy conservation the energy flux density of the
incident wave along the normal to the interface must be equal to
the sum of energy flux densities of the reflected waves. The
energy flux is a real quantity, and with our complex solutions it
is defined~\cite{kis} as
\begin{equation}\label{j1nn}
J_i=-\fr12\lt[\sigma^*_{il}\fr{du_l}{dt}+\sigma_{il}\fr{du^*_l}{dt}\rt]=-{\rm
Re}\lt[\sigma^*_{il}\fr{du_l}{dt}\rt],
\end{equation}
where Re($F$) means real part of $F$, and $*$ means complex
conjugation. After substitution of all the quantities we obtain
\begin{equation}\label{enr}
\J={\rm Re}(\rho\omega c_t^2u_0^2[E\A^*(\kk^*\cdot\A)+\kk^*]),
\end{equation}
where $E=1+\lambda/\mu$, and the amplitude $u_0$ of the
oscillations is taken into account in \eref{uu1}. Expression
\eref{enr} is valid for real and complex wave vectors and
polarizations. The absolute value of this flux for the incident
transverse wave is
\begin{equation}\label{enr1}
J=\rho\omega c_t^2u_0^2k.
\end{equation}
The flux of the LSW along the surface with account of its factor
$r^{(32)}$ is
\begin{equation}\label{enr2}
J_S^{3}(z)=\rho\omega u_0^2|r^{(32)}|^2c_l^2k_\|\exp(2K_lz).
\end{equation}
The ratio $Q=J_S^{(3)}(z=0)/J$ of two fluxes after substitution of
$r^{(32)}$ from \eref{q5} is
\begin{figure}[b]
{\par\centering\resizebox*{8cm}{!}{\includegraphics{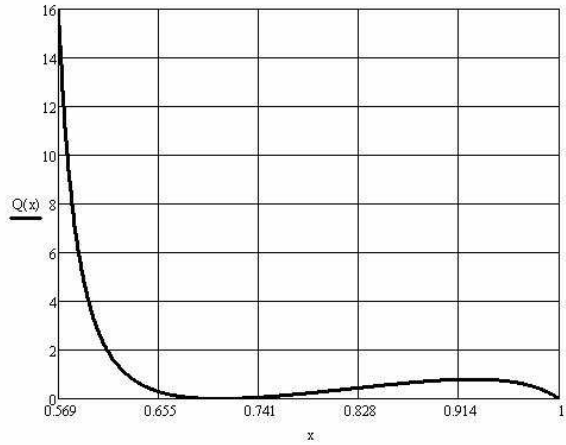}}\par}
\caption{\label{quak} Dependence of $Q(x)$ on $x=\cos\varphi$
calculated for $E\approx2.3$ like in stainless steel. The left
point on the abscise axis corresponds to critical
$\cos\varphi_c=c_t/c_l\approx0.57$, where $Q=Q_c=16$. In the case
of $E=1.2$ the similar calculations give
$\cos\varphi_c=c_t/c_l=0.674$ and $Q_c=323.616$.}
\end{figure}
\begin{equation}\label{enr3}
Q=\fr{k_\|}{k^{(2)}}\fr{c_l^2}{c_t^2}|r^{(23)}|^2=\fr{k_\|}{k^{(2)}}
\fr{16k^{(2)2}_{\bot}k_\|^2(k^{(2)2}-2k_\|^2)^2}{16K_l^2k^{(2)2}_{\bot}k_\|^4+(k^{(2)2}-2k_{\|}^2)^4},
\end{equation}
or\footnote{In~\cite{ifan} in this formula one factor
$\cos^2\varphi$ in denominator was missed}
\begin{equation}\label{enr4}
Q=
\fr{\cos\varphi\sin^2(4\varphi)}{4(\cos^2\varphi-\cos^2\varphi_c)\sin^2(2\varphi)\cos^2(\varphi)+\cos^4(2\varphi)},
\end{equation}
where $\varphi\equiv\varphi^{(2)}$. Dependence of this function on
$\cos\varphi$ is shown in Fig.~\ref{quak}. We see that the highest
energy density is accumulated in longitudinal surface wave, when
$\varphi$ is slightly less than $\varphi_c$. The higher is the
ratio $c_t/c_l$, the smaller is the critical grazing angle and the
higher is accumulation of the LSW energy density near the critical
angle. In the case of stainless steel the ratio $c_t/c_l$ is near
0.56, and $Q(\varphi_c)\approx16$.

\paragraph{Critical polarization.} Let's nevertheless find the
exact polarization of the total wave field on the interface, when
the grazing angle of the incident wave is $\varphi_c$. According
to \eref{eq10} the total displacement vector at $z=0$ is equal to
\begin{equation}\label{eq10a}
\A_t=\A^{(2)}+r^{(22)}(\varphi_c)\A^{(2)}_R+
r^{(32)}(\varphi_c)\A^{(3)}_R,
\end{equation}
where according to \eref{q5}
\begin{equation}\label{q5a}
r^{(32)}(\varphi_c)=2\sin(\varphi_c)\tan(2\varphi_c), \qquad
r^{(22)}(\varphi_c)=-1.
\end{equation}
Substitution of \eref{q5a} and
\begin{equation}\label{q5b}
\A^{(2)}=\ta \cos(\varphi_c)-\n \sin(\varphi_c),\qquad
\A^{(2)}_R=-[\ta \cos(\varphi_c)+\n \sin(\varphi_c)],\qquad
\A^{(3)}_R=\ta
\end{equation}
into \eref{eq10a} gives
\begin{equation}\label{eq10b}
\A_t=2\cos(\varphi_c)[\tan(\varphi_c)\tan(2\varphi_c)-1]\ta,
\end{equation}
where $\sin(\varphi_c)=c_t/c_l$.

From this expression it is seen, that at critical angle vibrations
are along the surface (such vibrations are the most devastating at
earthquakes), and their amplitude becomes especially large at
$\varphi_c\approx\pi/4$, i.e. in the case when $c_t/c_l\approx
0.7$. In the case of stainless steel with $c_t/c_l= 0.57$, though
$Q(\varphi_c)=16$ as is shown in Fig. \ref{quak}, the total
amplitude is only $\A_t(\varphi_c)=1.4\ta$. In the case of
$c_t/c_l= 0.65$ we get $\A_t(\varphi_c)=6.8\ta$ and
$Q(\varphi_c)=71$

At $\varphi<c_t/c_l$ polarization of LSW is directed along the
wave vector $\kk^{(3)}=(\kk_\|,iK_\bot)$, but this vector has an
imaginary normal component $K_\bot$. It means that vibrations in
LSW have also normal component, which is shifted by phase $\pi/2$
with respect to vibrations along the surface. However near the
critical angle the component $K_\bot$ is small and vibrations
along the normal can be neglected.

\section{An experiment with UCN to observe LSW}

\begin{figure}[h!]
{\par\centering\resizebox*{10cm}{!}{\includegraphics{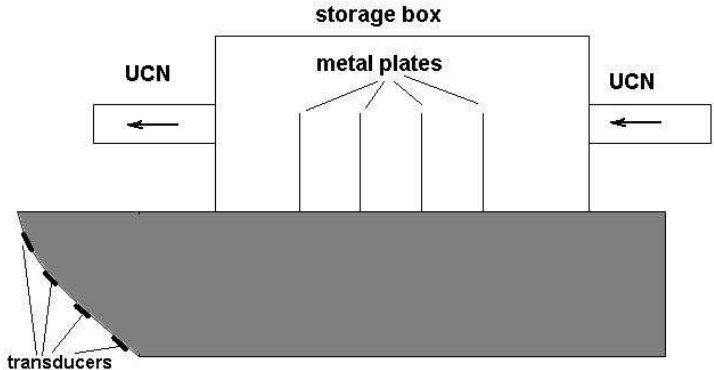}}\par}
\caption{\label{f2} Scheme of an experiment with UCN sensitive to
LSW.}
\end{figure}

Scheme of the proposed experiment with UCN is shown in Fig.
\ref{f2}. The floor of the storage box for UCN is a thick
basement. Several vertical thin plates are attached to it. And
transducers of shear waves with polarization in the vertical plane
generate ultrasound waves going to the storage floor at different
angles.
The UCN can be stored in the box or continuously flow through the
box, and dependence of the storage time or transmission of the box
on angle of incidence of the ultrasound wave is measured. If the
grazing angle of the incident ultrasound wave is close and
slightly below the critical one, the LSW created on the surface
will make oscillating the vertical plates, and the UCN in the box
will be heated at every collision with the plates and continuously
increase their energy up to the limit after which they will go
away through the box walls. It decreases the storage time or
transmission of the box.

For estimation of the effect one can consider some material like
stainless steel with density $\rho\approx8$ g/cm$^3$, the speed of
transverse waves $c_t\approx 3$ km/s and $c_t/c_l=0.57$, like that
used in Fig. \ref{quak}. Then ultrasound with $\omega=1$ MHz and
amplitude $u_0=0.1\,\mu$, has energy density $\rho\omega^2cu^2$ of
the the order of 10 W/cm$^2$. The speed $v_0$ of the horizontal
vibrations on the box floor surface at critical incidence angle
will be near $v_l=0.4$ m/s. It will create vibrations of the
vertical plates with the same speed. Every collision of UCN with
walls will increase in average their $v^2$ by $v_l^2=0.16$
m$^2$/s$^2$. Therefore, after 100 collisions with the walls their
energy will be high enough to penetrate the walls and to escape
from the storage box.

\begin{figure}[h!]
{\par\centering\resizebox*{10cm}{!}{\includegraphics{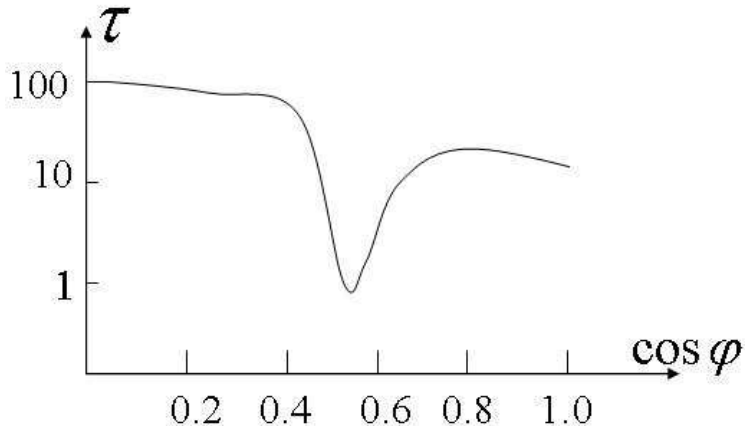}}\par}
\caption{\label{f3} Qualitative dependence of the storage time
$\tau$ on $\cos\varphi$, where $\varphi$ is the grazing angle of
the incident sound wave. The position of the dip depends on ratio
of $c_t/c_l$ of the material used for basement in Fig. \ref{f2}.
The depth and the width of the dip depend on the power and
frequency of the ultrasound.}
\end{figure}

The dependence of the storage time $\tau$ or transmission of the
box on cosine of the grazing angle $\varphi$ is expected to look
as shown in fig. \ref{f3}.
Observation of the dip will prove the appearance of the LSW and
therefore will prove the validity of the earthquake model.

\section{Conclusion}

The theory of LSW is presented. Though surface waves were studied
extensively~\cite{vic} the longitudinal one was
discovered~\cite{ifan} only because of  nonstandard approach to
theory of elasticity. The LSW appears at reflection from a free
surface of a shear wave with polarization in the incidence plane,
when grazing angle of the incident wave is below of the critical
one $\varphi_c=\arccos(c_t/c_l)$. Below and near $\varphi_c$ the
energy density of LSW can be orders of magnitude higher than
energy density of the incident wave. From the properties of LSW
one can conclude that they can have the same devastating effect as
earthquakes. Therefore LSW can be considered as a model of
earthquake waves. We described here an experiment, which can check
the properties of LSW. It is related to storage or transmission of
UCN. In the experiment one can see a sharp drop of storage time or
transmission of UCN through the storage box when LSW appears on
the surface of the storage box floor.

\section{History of submissions and rejections}

The paper was submitted to Phys.Rev.Lett. on 07.11. On 10.11 I
received a short letter from Senior Assistant editor with words
that my paper is not suitable for PRL. I submitted immediately to
Europhys.Lett and and this time the fate of the paper was
extraordinarily lucky. On 07.12 I received a positive referee
report, and the editor (it was Prof. Rudolf Treumann) returned me
my manuscript with completely corrected English. I cannot
withstand the desire to express my admiration and deep gratitude
to Prof. Treuman.

    In the new version, submitted to arXiv I added a paragraph
    at the end of section II.

\end{document}